# Airborne gamma-ray spectroscopy for modeling cosmic radiation and effective dose in the lower atmosphere

Marica Baldoncini[1,2], Matteo Albéri[1,2], Carlo Bottardi[1], Brian Minty[3], Kassandra G.C. Raptis[1,4], Virginia Strati[1,4], and Fabio Mantovani[1,2]

[1]Department of Physics and Earth Sciences, University of Ferrara, Via Saragat 1, 44121 - Ferrara, Italy
[2]INFN, Ferrara Section, Via Saragat 1, 44121 - Ferrara, Italy
[3]Minty Geophysics, PO Box 3299, Weston Creek, ACT, 2611, Australia
[4]INFN, Legnaro National Laboratories, Viale dell'Universitá, 2 - 35020 Legnaro (Padua), Italy

*Correspondence to:* Marica Baldoncini (baldoncini@fe.infn.it)

**Abstract.** In this paper we present the results of a ∼5 hour airborne gamma-ray survey carried out over the Tyrrhenian sea in which the height range (77-3066) m has been investigated. Gamma-ray spectroscopy measurements have been performed by using the AGRS_16L detector, a module of four 4L NaI(Tl) crystals. The experimental setup was mounted on the Radgyro, a prototype aircraft designed for multisensorial acquisitions in the field of proximal remote sensing. By acquiring high-statistics spectra over the sea (i.e. in the absence of signals having geological origin) and by spanning a wide spectrum of altitudes it has been possible to split the measured count rate into a constant aircraft component and a cosmic component exponentially increasing with increasing height. The monitoring of the count rate having pure cosmic origin in the >3 MeV energy region allowed to infer the background count rates in the $^{40}$K, $^{214}$Bi and $^{208}$Tl photopeaks, which need to be subtracted in processing airborne gamma-ray data in order to estimate the potassium, uranium and thorium abundances in the ground. Moreover, a calibration procedure has been carried out by implementing the CARI-6P and EXPACS dosimetry tools, according to which the annual cosmic effective dose to human population has been linearly related to the measured cosmic count rates.

## 1 Introduction

During the last few decades airborne gamma-ray spectroscopy (AGRS) has been demonstrated to be an extraordinarily powerful method for environmental monitoring, mineral exploration and geological mapping. Although far from being a novel technique, the frontiers of AGRS and its applications are continuously pushed forward thanks to advances in multichannel processing, statistical methods for spatial resolution enhancement and data analysis procedures (Beamish, 2016; Guastaldi et al., 2013; Minty and Brodie, 2016; Strati et al., 2015; Sanderson et al., 2008).

The improvement of AGRS data quality evolved side by side with the integration of geological data via increasingly refined statistical and geostatistical methods: the combined effect of both aspects allowed to go beyond traditional mineral exploration and lead to the investigation of new multidisciplinary fields, like landslide monitoring (Baroň et al., 2013), peat thickness estimation (Keaney et al., 2013), prediction models for trees' growth (Mohamedou et al., 2014), radioelement distribution in weathered materials (Iza et al., 2016) and precision agriculture (Söderström et al., 2016). Concurrently, the potentialities of the AGRS technique in the sector of homeland security have been widely explored in terms of feasibility of real-time identification of anthropogenic radionuclides on top of the natural background (Cardarelli et al., 2015; Detwiler et al., 2015; Kock et al., 2012; Kulisek et al., 2015; Kock et al., 2014) and of merging and comparing results from multi-regional AGRS campaigns performed in the framework of intercomparison exercises (Bucher et al., 2009). In the light of environmental contamination assessment, the detection of anthropogenic radionuclides emitting low energy gamma-rays (e.g. $^{137}$Cs and $^{131}$I) together with the employment of new unmanned aerial vehicle (UAV) devices, characterized by different detection performances compared to standard acquisition systems, are reawakening the effort in estimating detectors efficiencies and minimum detectable activities (MDA) (Cao et al., 2015; Gong et al., 2014; Tang et al., 2016).

In order to address the AGRS new challenges, an adequate understanding and knowledge of the background spectral components is mandatory for processing airborne gamma-ray spectrometric data. Indeed, independently from the specific application, from the employed aircraft and from the particular radionuclides under investigation, cosmic radiation is an ever-present component: the better it is characterized, the easier its identification would be. In this context, this paper provides insights which are significative for multiple disciplines due to the diagonal nature of the topic of cosmic radiation in the environment.

Galactic cosmic ray particles, with energies extending up to few $10^{20}$ eV (Bird et al., 1993; Pierre Auger Collaboration, 2010), are produced outside the Solar System and are constituted by a nucleonic component (98%) and electrons (2%). The nucleonic component is primarily made up by protons (∼85% of the flux) and alpha particles (∼12%), with a remaining fraction comprising heavier nuclei (UNSCEAR, 2008). In entering the Earth's atmosphere, these particles collide with atoms of air constituents, giving rise to cascades of secondaries, including neutrons, pions, muons and gamma radiation. In airborne gamma-ray spectroscopy counts collected in the range 3-7 MeV allow to identify the gamma component of cosmic rays, as the end point of gamma-rays of terrestrial origin corresponds to the $^{208}$Tl emission at 2.614 MeV. This information can be used not only for predicting the cosmic background in the $^{40}$K, $^{214}$Bi and $^{208}$Tl photopeak energy windows, but also for assessing the cosmic radiation dose to the human population.

The annual effective dose rate due to cosmic ray exposure averaged over the world's population has been estimated to be 0.38 mSv/y by UNSCEAR (2008), although recent efforts have been done in order to give more accurate evaluations on the base of advanced cosmic-ray fluxes calculation and refined grid databases of population and terrain elevation models (Sato, 2016). These estimations take into account the amount of time people spend indoor (80% of the day) and the mean thickness of the walls acting as a shield for the cosmic radiation. Cosmic dosimetric measurements are generally focused on the assessment of air crew members exposure. There are also regional measurement campaigns addressing the question of outdoor population dose exposure (Cavalcante et al., 2011; Strand et al., 2002; Tsui et al., 1991). In this context, the calibration of an airborne gamma-ray detector for the assessment of cosmic dose rates can provide a supplementary technique for the cosmic exposure assessment with respect to in-situ measurements.

In this work we present the results of a ∼5 hours AGRS survey over the sea dedicated to the measurement of the gamma radiation originating from the aircraft materials and cosmic rays, which constitute a background source for the estimation of the gamma radiation of terrestrial origin coming from $^{40}$K, $^{214}$Bi (eU) and $^{208}$Tl (eTh). The AGRS non-geological background radiation is investigated with 17612 1 second measurements in a wide range of elevations (77-3066) m. The acquisition of spectra over water at a number of different heights indeed provides a way to split the constant contribution coming from the radioactivity of the aircraft from the height dependent contributions associated with cosmic radiation and, if present, with atmospheric radon (Minty et al., 1997). Moreover, we study a linear calibration curve which allows to convert count rates into the electromagnetic shower component of the cosmic effective dose (CED$^{EMS}$) based on two cosmic ray dosimetry software tools: CARI-6P (FAA, 2014) and Excel-based program for calculating atmospheric cosmic ray spectrum (EXPACS) (Sato, 2015). A procedure for the calculation of cosmic effective dose to human population (CED) is finally proposed.

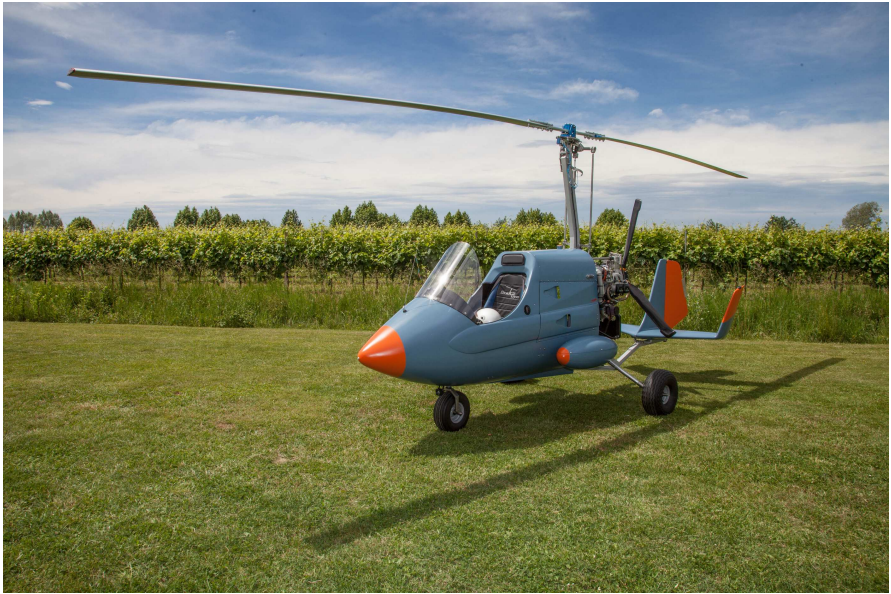

**Figure 1.** Picture of the Radgyro, the autogyro dedicated to multispectral airborne acquisitions, used for the AGRS surveys over the sea.

## 2 Instruments and methods

### 2.1 Experimental setup, survey and data

Airborne gamma-ray surveys over the sea were carried out with the Radgyro (Fig. 1), a prototype aircraft dedicated to multispectral airborne measurements. The Radgyro is an autogyro specifically designed to host a large network of sensors which are able to investigate simultaneously and independently the electromagnetic spectrum in a variety of spectral ranges, from thermal infrared (13000-7500 nm) to gamma radiation ($3 \cdot 10^{-3}$ - $4 \cdot 10^{-4}$ nm). The high autonomy and payload (3 hours for a 120 kg equipment weight), the modularity of the acquisition system, together with the possibility of time correlating the information coming from the different sensors, make the Radgyro a unique aircraft in the field of proximal remote sensing.

During the surveys the Radgyro position is recorded every second by a GPS system composed of 2 u-blox EVK-6T antennas (Albéri et al., 2017). Gamma radiation is measured with the AGRS_16L system, a modular scintillation detector hosted in the central region of the aircraft hull and composed of four 10 cm × 10 cm × 40 cm NaI(Tl) crystals, for a total detection volume of 16L. Each detector has a 1 mm thick stainless steel shielding and is coupled with a PMT base which receives the voltage supply from a power unit shared among all the sensors mounted on the aircraft. The signals are acquired in list mode (event by event) by a CAEN DT5740 module, a 32 channel 12 bit 62.5 MS/s waveform digitizer.

AGRS raw data are acquired event by event separately for each of the four NaI(Tl) detectors: each list mode file contains the time stamp of a given energy deposition (in units of digitizer clock) together with the corresponding acquisition channel. The list mode files are cut offline for each detector in order to produce 1 second acquisition spectra which subsequently undergo an energy calibration procedure. The latter is performed by determining with a Gaussian fit the positions of the prominent $^{40}$K and $^{208}$Tl photopeaks in 600 seconds spectra acquired on the ground before the take off. A linear function is then fitted to the photopeaks' positions for estimating the energy corresponding to the first acquisition channel (keV) and the gain (keV/channel). Summing up the four calibrated spectra it is possible to obtain the gamma-ray 1 second spectrum acquired by the entire 16L detection volume, which has an energy end point of 7 MeV.

For what concerns the Radgyro positioning, each GPS antenna produces two separate files, one containing the temporal information in terms of PC and GPS acquisition times, the second is a binary file which is processed with the goGPS software (Herrera et al., 2016) for the extraction of the standard NMEA sentence. The mean 1 second position and altitude above sea level of the Radgyro is computed as the average of the coordinates obtained from the single GPS receivers. As both the radiometric and positioning data are acquired with the same PC, the computer time stamp is used for the synchronization of the different devices (Albéri et al., 2017).

Airborne gamma-ray background calibration surveys were performed in a series of 4 flights over the Tyrrhenian Sea close to Viareggio (Tuscany, Italy) with typical horizontal and vertical velocities of ∼20 m/s and ∼0.8 m/s, respectively. In order to avoid taking into account gamma-ray signals potentially spoiled by ground radiation, gamma-ray measurements acquired at a distance from the coast less than 300 m have been excluded from the analysis. In Fig. 2 the effective paths of the different flights are shown, which correspond to a total acquisition time of 17612 seconds and an explored range of altitudes going from 77 m to 3066 m. In Table 1 a summary of the main parameters related to each of the 4 flights is shown.

According to the purpose of the experiment, the flight paths have been planned with the aim of investigating the entire reported range of heights with enough statistics for well constraining the analysis of the altitude dependent gamma-ray cosmic component. This strategy, together with the flight conditions and the non feasibility for the Radgyro to hover at a given elevation, allowed us to collect the elevation flight statistics shown in Fig. 3.

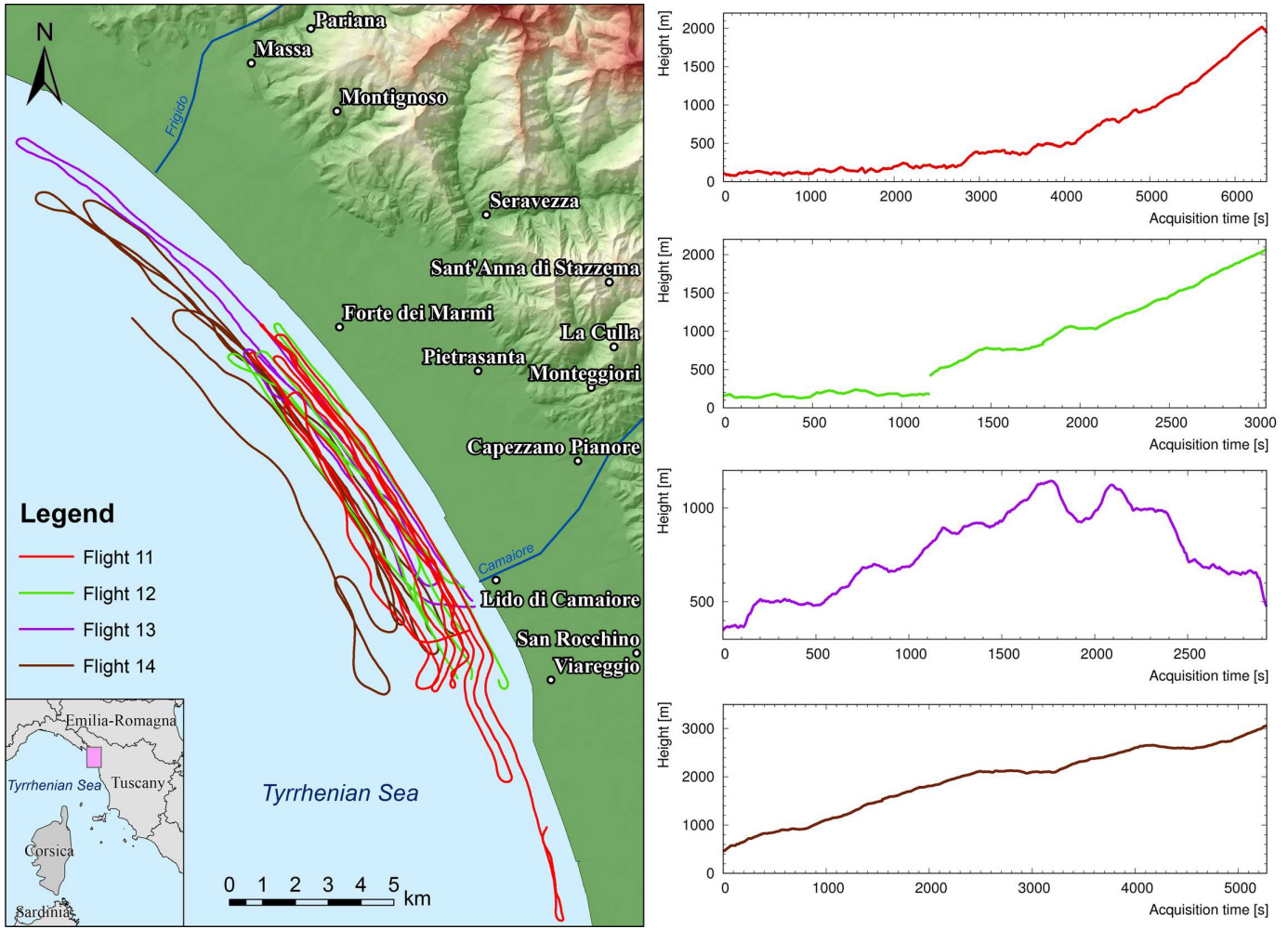

**Figure 2.** The left panel shows a map of the effective flight lines of the surveys over the sea performed near Viareggio (Tuscany, Italy). The acquisition tracks are the ones corresponding to data points acquired at a minimum distance from the coast of 300 m. The four panels on the right show the altitude profiles for the different flights. On the x axis the effective acquisition time for each individual flight is reported (see Table 1).

**Table 1.** Summary of the main parameters for each of the 4 surveys over the sea. For each flight the ID, date, time, minimum and maximum altitude and acquisition time are reported, respectively. In the case of flights 11 and 14, 83 seconds and 30 seconds have been cut due to some radiofrequency interference between the PMT and the aircraft transponder. The long interruption of the data taking of flight 12 (2531 seconds) has been imposed by civil traffic of the Pisa airport.

| Flight ID | Date | Time | z min [m] | z max [m] | Acquisition time [s] |
|---|---|---|---|---|---|
| 11 | 30/03/2016 | 17:42:10<br>19:29:43 | 77 | 2019 | 6370 |
| 12 | 31/03/2016 | 18:13:55<br>19:46:47 | 126 | 2070 | 3041 |
| 13 | 05/04/2016 | 11:39:53<br>12:28:36 | 348 | 1144 | 2924 |
| 14 | 05/04/2016 | 16:37:16<br>18:05:43 | 461 | 3066 | 5277 |
| Global | | | 77 | 3066 | 17612 |

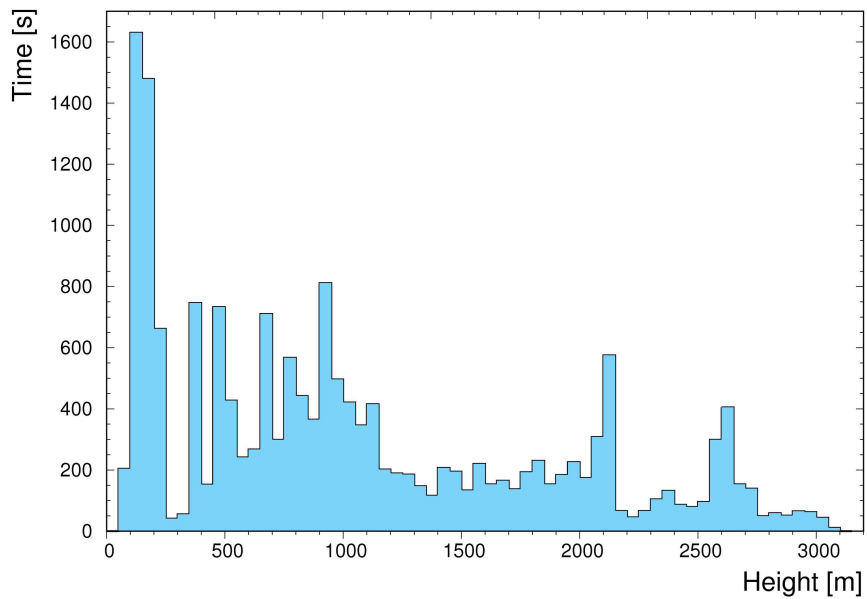

**Figure 3.** Histogram describing the effective overall temporal statistics: the data taking time at a given survey altitude is shown, with an altitude binning of 50 m.

## 3 Theoretical modeling and data analysis

Airborne gamma-ray spectroscopy measurements are affected by background radiation, which can be considered as radiation not originating from the Earth's surface and which has to be removed during data processing. The three major sources of background radiation are cosmic background, instrumental plus aircraft background and atmospheric radon ($^{222}$Rn).

The cosmic gamma background resulting from the interaction of cosmic secondary radiation interaction with the air, the aircraft and the detector materials is foreseen to monotonically increase with increasing altitude. Concerning the energy dependence, the cosmic-induced gamma-ray energy spectrum is expected to have a polynomial dependence with respect to gamma-ray energy (Sandness et al., 2009). The count rate (CR) energy dependence of the cosmic component is reconstructed according to a polynomial function having the following expression:

$$CR(E) = aE^b + c \qquad (1)$$

where $E$ is the gamma-energy in MeV and $a$, $b$ and $c$ are constants for a spectrum measured at a given altitude. The energy dependence of the CR has been estimated by fitting the measured spectrum with the above model function both in the 0.8-7 MeV energy range and in the 3-7 MeV energy range, called respectively the Full Energy Window (FEW) and the Cosmic Energy Window (CEW). A third fit has been performed using as input data points the measured CRs in the CEW, plus the three points corresponding to the estimated CRs due to cosmic radiation in the $^{40}$K, $^{214}$Bi and $^{208}$Tl photopeak energy windows (see Table 3), which have been determined on the base of the linear regression parameters reported in Table 5.

In Table 2 the results of this analysis in two different ranges of altitudes is reported. In both cases radiometric data have been acquired above 2000 m, where the presence of atmospheric radon is negligible (see Section 4). Fig. 4 shows an example of background airborne gamma-ray spectrum measured with the AGRS_16L together with the three curves resulting from the different fitting procedures. From this exercise it is possible to evince that the fitting of the measured spectrum is dependent on the energy range chosen, as the spectral shape under reconstruction contains different pieces of information in the CEW and in the FEW. Using only the CEW for constraining the cosmic spectral shape from one side assures the pure cosmic nature of the counting statistics, but on the other side the sole reconstruction of the spectral high energy tail prevents a correct estimation of the curve slope in the low energy range as emphasized by the large uncertainties on the best fit parameters. By fitting in the FEW the steep behavior at low energies is reproduced: however in this case the measurement under reconstruction contains not only the cosmic contribution to the signal, but also the signal coming from the equipment radioactivity and in particular from the $^{40}$K, $^{214}$Bi and $^{208}$Tl decay series. On the other hand, the idea behind the third fitting approach is to reinforce the fit performed by using the sole count rates in the CEW with the addition of three relatively well separated points corresponding to the cosmic CRs in the $^{40}$K, $^{214}$Bi and $^{208}$Tl photopeak energy windows. Among the above mentioned three strategies this is the one providing the most reliable estimation of the cosmic spectral shape in the FEW.

Instrumental and aircraft background correspond to the constant gamma signal generated by trace amounts of K, U and Th contained in the detector materials and ancillary equipment, together with the aircraft material itself. $^{222}$Rn, the only gaseous daughter product of the $^{238}$U decay chain, can escape from rocks and soils and, considering its 3.8 days half-life, can accumulate in the lower atmosphere. Its gamma-emitting daughter nuclei $^{214}$Bi and $^{214}$Pb can attach to airborne aerosols and dust particles, giving rise to the atmospheric radon background gamma signal (Grasty and Minty, 1995). The determination of the K, U and Th ground concentrations during an airborne gamma-ray survey relies on the estimation of the background corrected CRs recorded in the $^{40}$K, $^{214}$Bi (eU) and $^{208}$Tl (eTh) photopeak energy windows, called KEW, BEW and TEW, respectively (see Table 3).

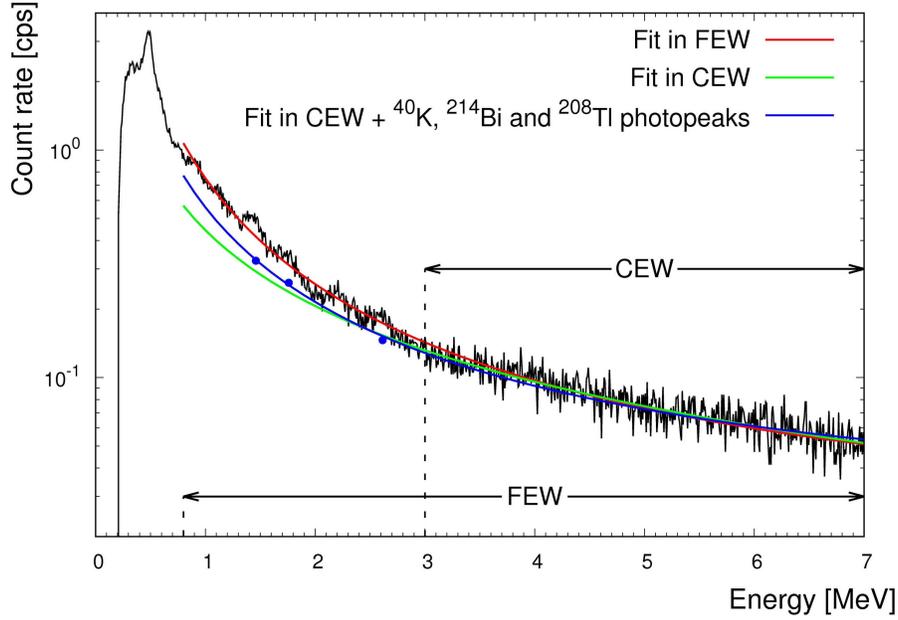

**Figure 4.** Gamma-ray spectrum composed of 870 1 second spectra acquired in the elevation range 2050-2150 m (black solid line). The red solid line shows the fitting curve obtained using as model function Eq. 1 and as energy fitting range the FEW, the green solid line shows the curve obtained by fitting the measured spectrum in the CEW. The blue points correspond to the CRs in the KEW, BEW and TEW associated with the cosmic induced background and obtained on the base of the linear relation having as parameters the ones reported in Table 5. The blue solid line is the result of the fit of the measured spectrum in the CEW and of the blue points.

**Table 2.** Fit parameters of the CR energy dependence modeled with Eq. 1 for two spectra measured at 2100 m and 2650 m for respectively 870 seconds and 550 seconds. For each measured spectrum the fit has been performed in the FEW, in the CEW and in the CEW plus the $^{40}$K, $^{214}$Bi and $^{208}$Tl photopeaks.

| z range [m] | Fit energy range | (a $\pm$ $\delta$a) [cps] | b $\pm$ $\delta$b | (c $\pm$ $\delta$c) [cps] |
|---|---|---|---|---|
| | FEW | 0.73 $\pm$ 0.10 | -1.62 $\pm$ 0.40 | 0.02 $\pm$ 0.03 |
| 2050 - 2150 | CEW | 0.44 $\pm$ 0.42 | -1.11 $\pm$ 1.60 | 0.00 $\pm$ 1.40 |
| | CEW + $^{40}$K, $^{214}$Bi and $^{208}$Tl photopeaks | 0.54 $\pm$ 0.04 | -1.49 $\pm$ 0.05 | 0.02 $\pm$ 0.01 |
| | FEW | 0.90 $\pm$ 0.11 | -1.53 $\pm$ 0.33 | 0.02 $\pm$ 0.04 |
| 2600 - 2700 | CEW | 0.62 $\pm$ 0.61 | -1.14 $\pm$ 1.66 | 0.00 $\pm$ 1.87 |
| | CEW + $^{40}$K, $^{214}$Bi and $^{208}$Tl photopeaks | 0.71 $\pm$ 0.05 | -1.45 $\pm$ 0.03 | 0.02 $\pm$ 0.01 |

**Table 3.** Energy windows for natural and cosmic radiation used for the background calibration of the AGRS_16L system. The last two columns report for each energy window the measured CR for gamma-ray spectra acquired at the altitude range 2050-2150 m and 2600-2700 m, respectively.

| Energy Window | Emission line [keV] | Energy range [keV] | Measured CR [cps] (2050 - 2150) m | Measured CR [cps] (2600 - 2700) m |
|---|---|---|---|---|
| KEW | 1460 | 1370 - 1570 | 12.2 | 15.0 |
| BEW | 1765 | 1660 - 1860 | 8.7 | 11.1 |
| TEW | 2614 | 2410 - 2810 | 8.8 | 11.9 |
| CEW | / | 3000 - 7000 | 41.9 | 54.8 |

Aircraft and cosmic background calibration flights are usually performed offshore for a typical altitudes range of (1500 - 3000) m above the ground level in order to avoid the contamination from terrestrial radiation and radon decay products (International Atomic Energy Agency, 1991). In this scenario, as the instrumental background is supposed to be constant and the gamma cosmic background is expected to exponentially increase with increasing height, the measured CRs in the $i'th$ energy window during a calibration flight over the sea is predicted to follow the subsequent equation:

$$n^i(z) = A^i e^{\mu^i z} + B^i \tag{2}$$

where $n^i$ is the CR in the $i'th$ energy window (with $i$ = KEW, BEW, TEW and CEW) $A^i$, $\mu^i$ and $B^i$ are constants (International Atomic Energy Agency, 1991; Minty et al., 1997).

The CR in the natural radionuclides energy windows are expected to be linearly related to the count rate in the CEW, as stated in the following equation:

$$n^i = a^i + b^i n^{\text{CEW}} \tag{3}$$

where $n^i$ is the CR in the $i'th$ energy window (with $i$ = KEW, BEW, TEW), $a^i$ is the aircraft background CR in the $i'th$ energy window, $b^i$ is the cosmic stripping ratio (i.e. the cosmic background CR in the $i'th$ energy window normalized to unit counts in the CEW) and n$^{\text{CEW}}$ is the CR in the CEW. The parameter $a^i$ is the expected CR for null cosmic CR and therefore represents the constant background component generated by the Radgyro and by the detectors materials. Determining these linear functions for the natural radionuclides energy windows allows to correct the CRs measured at a given height during regional AGRS surveys for the aircraft and height dependent cosmic ray backgrounds, provided the monitoring of the CR in the CEW.

Eq. 2, as well as Eq. 3, holds in the absence of any terrestrial and atmospheric radon radiation component. A potential radon contamination in any case would act on the CRs in the KEW and BEW but not on the CRs in the TEW and CEW as they are not affected by the lower energy gamma emissions of radon daughter nuclei. The presence of a radon background component in the measured CRs can be generally identified as a breakdown of the linear relationship between the cosmic and the $^{214}$Bi CRs at low elevations (Baldoncini et al., 2017). The estimated CRs in the energy windows of interest have been clustered according to an altitude binning of 15 m, which is conservative with respect to the estimated accuracy on the vertical position resulting from the combination of all the altimeters present on board of the Radgyro (Albéri et al., 2017). The CRs used as input for the background modeling are therefore estimated summing all the input CRs acquired in the same elevation bin and dividing by the number of 1 second spectra entering the summation.

The parameters of the exponential curves $A^i$, $\mu^i$ and $B^i$ have been determined via the minimization of the $\chi^2$ function:

$$\chi^2_{\text{exp}} = \sum_{j=1}^{\text{nbin}} \left[ \frac{n_j^i - \left(A^i e^{\mu^i z_j} + B^i\right)}{\sigma_{n_j^i}} \right]^2 \qquad (4)$$

where nbin is equal to the number of elevation bins entering the $\chi^2$ minimization, $n_j^i$ is the average CR obtained for the $j'th$ elevation bin in the $i'th$ energy window, $z_j$ is the average elevation obtained for the $j'th$ elevation bin and $\sigma_{n_j^i}$ is the 1 sigma uncertainty associated to the counting statistics, corresponding to the square root of the total counts recorded at $z_j$ in the $i'th$ energy window divided by the acquisition time at $z_j$.

The objective $\chi^2$ function to be minimized for determining the linear curve parameters has instead to be built taking into account not only the statistical error associated to the quantity $n^i$ but also the uncertainty on the "independent variable" $n^{\text{CEW}}$. Therefore, the adopted definition for the $\chi^2$ function is:

$$\chi^2_{\text{lin}} = \sum_{j=1}^{\text{nbin}} \frac{\left[n_j^i - \left(a^i + b^i n_j^{\text{CEW}}\right)\right]^2}{\left(\sigma_{n_j^i}\right)^2 + \left(b^i \sigma_{n_j^{\text{CEW}}}\right)^2} \qquad (5)$$

Monitoring the CEW in principle can be used for estimating the CED to human population. Gamma-ray spectrometers for dosimetric measurements are generally calibrated by exposing them to certified radiation fields, which can be collimated beams at irradiation facilities, calibrated radioactive point sources with known activities covering both high and low energy ranges or calibration pads generally made of concrete and doped with radionuclides of known gamma dose-rates (Grasty et al., 2001; Tuan et al., 2013; Mercier and Falguéres, 2007).

In the last decades various codes devoted to the calculation of the aircraft crew's exposure to cosmic radiation have been developed on the base of Monte Carlo techniques, analytical solutions and empirical data fitting (Spurný and Daschev, 2002; Kleinschmidt and Watson, 2016; Bottollier-Depois et al., 2009). Since most of them are user friendly and well tested, their adoption for the calibration of an AGRS detector for cosmic dose estimation can be a valid option with respect to traditional characterization procedures. The popular software CARI-6P allows to calculate the different components of the cosmic effective dose received by an individual at typical cruise altitudes by relying on analytic calculations of particle transport through the atmosphere FAA (2014). The EXPACS dosimetry tool permits to model the fluxes of different cosmic particles in the lower atmosphere thanks to air shower simulation performed by Particle and Heavy Ion Transport code System (PHITS)(Sato, 2015).

**Table 4.** Fit parameters of the model curve formulated by Eq. 2 describing the CR dependence with respect to the elevation for the CRs measured in the TEW and in the CEW. The last column reports the value of the reduced $\chi^2$ obtained at the end of the minimization procedure.

| Energy Window | (A $\pm$ $\delta$A) [cps] | ($\mu \pm \delta\mu$) [m$^{-1}$] | (B $\pm$ $\delta$B) [cps] | Reduced $\chi^2$ |
|---|---|---|---|---|
| TEW | 2.4 $\pm$ 0.2 | (5.5 $\pm$ 0.2) $\cdot 10^{-4}$ | 1.6 $\pm$ 0.2 | 0.94 |
| CEW | 11.4 $\pm$ 0.3 | (5.9 $\pm$ 0.1) $\cdot 10^{-4}$ | 2.0 $\pm$ 0.4 | 1.12 |

Both codes require information on the altitude, the geographic location and the time period, the latter related to changes in the Earth's magnetic field and in solar activity. Since the count rate in the CEW measured by a gamma spectrometer during a calibration flight is related to the electromagnetic shower (CED$^{EMS}$), knowing the temporal and spatial coordinates of the survey it is possible to characterize a calibration curve, which depends on the detector and on the dosimetry software tool. Once the calibration parameters have been calculated, subsequent AGRS acquisitions can provide a direct experimental measurement of the CED$^{EMS}$, which can be checked a posteriori with the estimation given by the dosimetry code.

## 4 Results and Discussion

In this section we report the results regarding the background calibration of the AGRS_16L spectrometer performed via the analysis of 1 second gamma ray spectra acquired during a 17612 seconds airborne survey over the sea. For $^{40}$K and $^{214}$Bi the relation between $n^i$ and the altitude above the sea level is not guaranteed to be purely exponential down to low elevations, as the CRs in the $^{40}$K and $^{214}$Bi photopeaks may be contaminated by the presence of atmospheric radon. As already mentioned, this potential contamination also translates in a deviation from a purely linear relation between $n^i$ and n$^{CEW}$ at low elevations. The concentration of $^{222}$Rn in the atmosphere can change considerably according to the different diffusion conditions. Nevertheless, above 1000-1500 m, mean $^{222}$Rn concentrations of the daytime atmosphere drop sharply to values compatible with zero (around 2 $\pm$ 2 Bq/m$^3$) and then slowly reduce further with height until they reach 0.3 $\pm$ 0.4 Bq/m$^3$ above 3000 m (Williams et al., 2010). In our analysis the CRs in the KEW and in the BEW is conservatively studied only for altitudes greater than 2000 m.

Fig. 5a shows the experimental CRs in the CEW, distinguished by colour according to the different flight IDs: the homogeneity of this partial datasets assures that there are no systematic effects related to the different acquisition times. Fig. 5b shows the experimental data for the CRs in the CEW obtained from the entire dataset, with the superimposed curve resulting from the minimization of the $\chi^2$ function described by Eq. 4. The values of the fitting curve parameters are reported in Table 4.

The 1.12 reduced chi-square value denotes a good agreement between the model function and the experimental data. Although the parameters $A$ and $B$ in the CEW (Table 4) are affected by uncertainties having different order of magnitudes, at the nominal 100 m survey height of an airborne survey the two uncertainties separately produce approximately the same variation on the estimated CRs, which is below 3%. Thanks to the high acquisition statistics and to the wide range of investigated altitudes, the fit well constrains the value of the $\mu$ parameter entering the exponential dependency, which is estimated with an uncertainty of 2%.

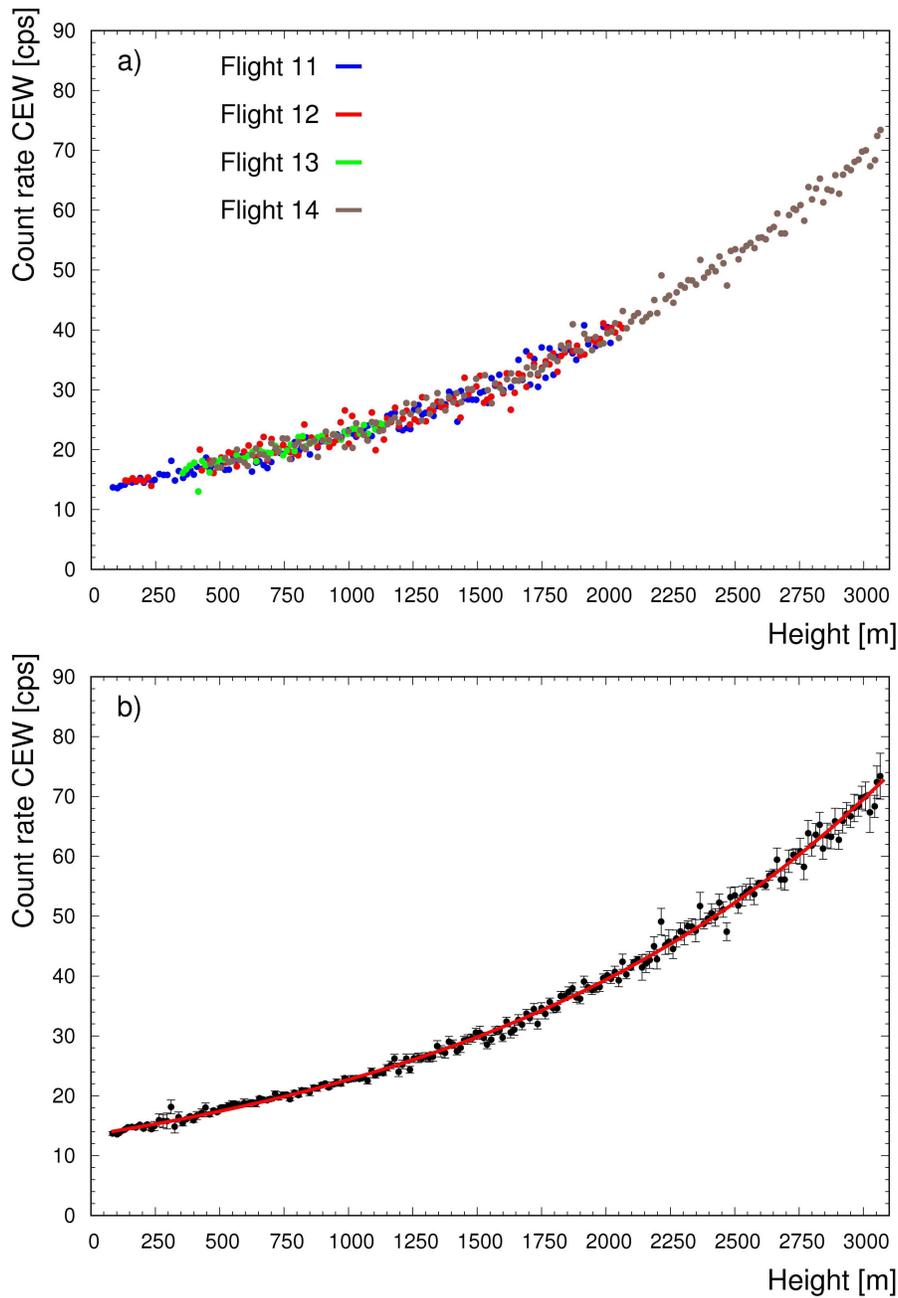

**Figure 5.** Panel a) displays the CR in the CEW as function of the altitude for the four different flights carried out during the background calibration survey over the sea. Data from different flights sit on top of each other, excluding systematic effects associated to the different acquisition times. Panel b) shows the CR in the CEW obtained from the entire dataset (black points) as function of the altitude with the superimposed exponential fit function (red solid line). Each point populating the global dataset has been obtained by clustering with an altitude binning of 15 m the spectra measured in that specific height range, disregarding any flight ID classification.

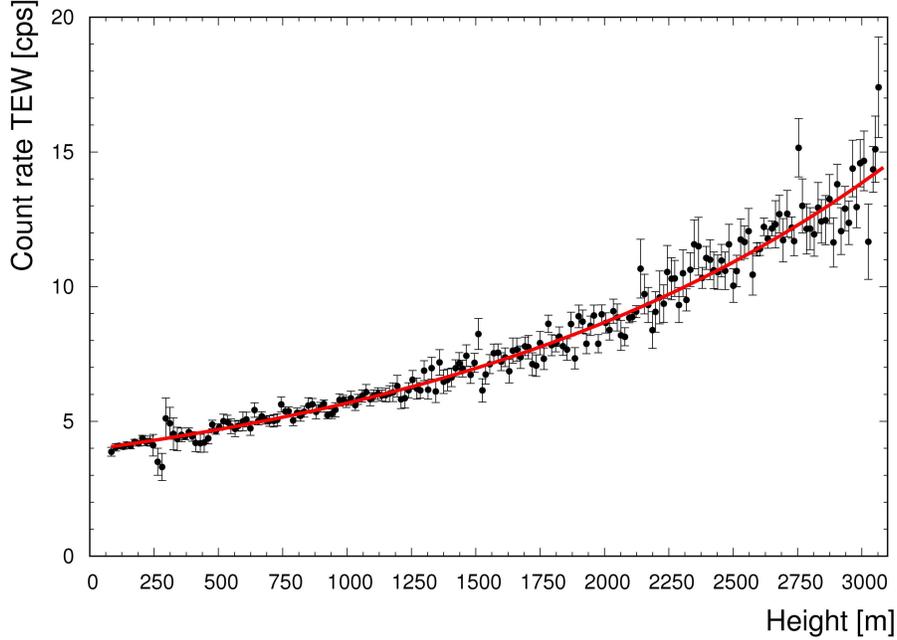

**Figure 6.** Plot of the experimental CR in the TEW as function of the altitude (black points) together with the corresponding fitting curve (solid red line).

Fig. 6 shows the experimental CRs in the TEW evaluated on the entire dataset, together with the best fit exponential curve, whose parameters values are listed in Table 4. Also in this case the reduced chi-square value reflects the high data quality as well as the goodness of the model function in interpreting the measured CRs. The impact of the fitting parameter uncertainties on the estimated CR is negligible for what concerns $\mu$ while the uncertainties on $A$ and $B$ in the TEW individually give rise to a 5% variation of the predicted CR at 100 m.

For both the CEW and the TEW, the minimization of the $\chi^2$ functions defined by Eq. 4 has been performed over the whole altitude range, corresponding to 200 height bins having a 15 m width. In both cases it is possible to recognize the presence of high statistics experimental points for height values below 200 m and around approximately 900, 2100 and 2650 m, which reflect the time flight statistics illustrated in Fig. 3. As a result of the definition of the objective $\chi^2$ function, the discrepancy between the fitting function and the data is minimum in correspondence of the experimental points having the smallest statistical uncertainty.

In International Atomic Energy Agency (1991) and Grasty and Minty (1995) an analogous study of the CR in the TEW as function of altitude is shown: this kind of reconstruction is carried out in both cases with a NaI spectrometer having 33.6 L volume, which precludes the possibility of a direct comparison with the results of this study. However, from a qualitative point of view, it emerges that the $\mu$ coefficient entering the exponential dependence (and essentially quantifying the rate of increase of the counting statistics) is for the three cases in the range $(4 - 6) \cdot 10^{-4}$ m$^{-1}$. Previous studies focused on a different altitude range, from around 1500 m to 4500 m: in this framework, this work demonstrates that the CR both in the CEW and in the TEW maintains its exponential behavior down to tens of meters above sea level.

**Table 5.** Fit parameters of the model curve formulated by Eq. 3 describing the dependence of the count rates in the KEW, BEW and TEW with respect to the CR in the CEW. The last column reports the value of the reduced $\chi^2$ obtained at the end of the minimization procedure.

| Energy Window | (a $\pm$ $\delta$a) [cps] | (b $\pm$ $\delta$b) [cps/cps in CEW] | Reduced $\chi^2$ |
|---|---|---|---|
| KEW | $3.7 \pm 0.4$ | $0.20 \pm 0.01$ | 1.00 |
| BEW | $2.0 \pm 0.4$ | $0.16 \pm 0.01$ | 1.02 |
| TEW | $1.58 \pm 0.04$ | $0.179 \pm 0.002$ | 1.02 |

The analysis of the exponential trend of the CRs with respect to the altitude could have been done in principle also for the CRs in the KEW and in the BEW, restricting the fitting domain to the range of altitudes greater than 2000 m. However, as the slope of the CR increase with respect to the altitude is small in the 2000 m to 3000 m height domain, fitting in the 2000 - 3000 m height domain would suffer the lack of the low altitude tail, producing incorrect extrapolations down to sea level. This point can be a trigger for a deeper investigation, as it can potentially be a way for exploring the content of $^{222}$Rn in the lower atmosphere (Baldoncini et al., 2017).

Fig. 7 shows the experimental data with the superimposed linear curve resulting from the minimization of the $\chi^2$ function described by Eq. 5, where the number of bins is equal to 200 for the TEW and is equal to 72 for the KEW and the BEW. Table 5 lists the fitting parameters together with the associated uncertainties and the reduced $\chi^2$ value, which is almost 1 for all the three energy windows. In the perspective of using the linear relations for applying the Window Analysis Method (International Atomi Energy Agency, 2003) to airborne gamma-ray spectra, the uncertainties estimated in Table 5 are relevant for attempting an evaluation of systematics associated with aircraft and cosmic background corrections. With the hypothesis of flying at 100 m height, the mentioned background CR is $(6.5 \pm 0.5)$ cps in the KEW, $(4.3 \pm 0.6)$ cps in the BEW and $(4.1 \pm 0.1)$ cps in the TEW.

For the CR in the TEW, as both the exponential and linear curve reconstructions have been performed, it is possible to check the consistency of the obtained results according to the existing relations among the fit parameters. On the base of the expected value of the CRs in the CEW and in the TEW at zero altitude, it is also possible to establish the following relationship among fit parameters:

$$A^{TEW} + B^{TEW} = a^{TEW} + b^{TEW} \left( A^{CEW} + B^{CEW} \right) \tag{6}$$

Adopting the parameters reported in Table 4 one can calculate the left hand side of Eq. 6, which corresponds to $(4.0 \pm 0.4)$ cps. The right hand side of the equation can be estimated using the parameters listed in Table 4 and in Table 5, which provide a count rate of (4.0 ± 0.2) cps. The perfect agreement gathered from this analysis is an important internal consistency check of the goodness of both the exponential and linear model function in interpreting the experimental data.

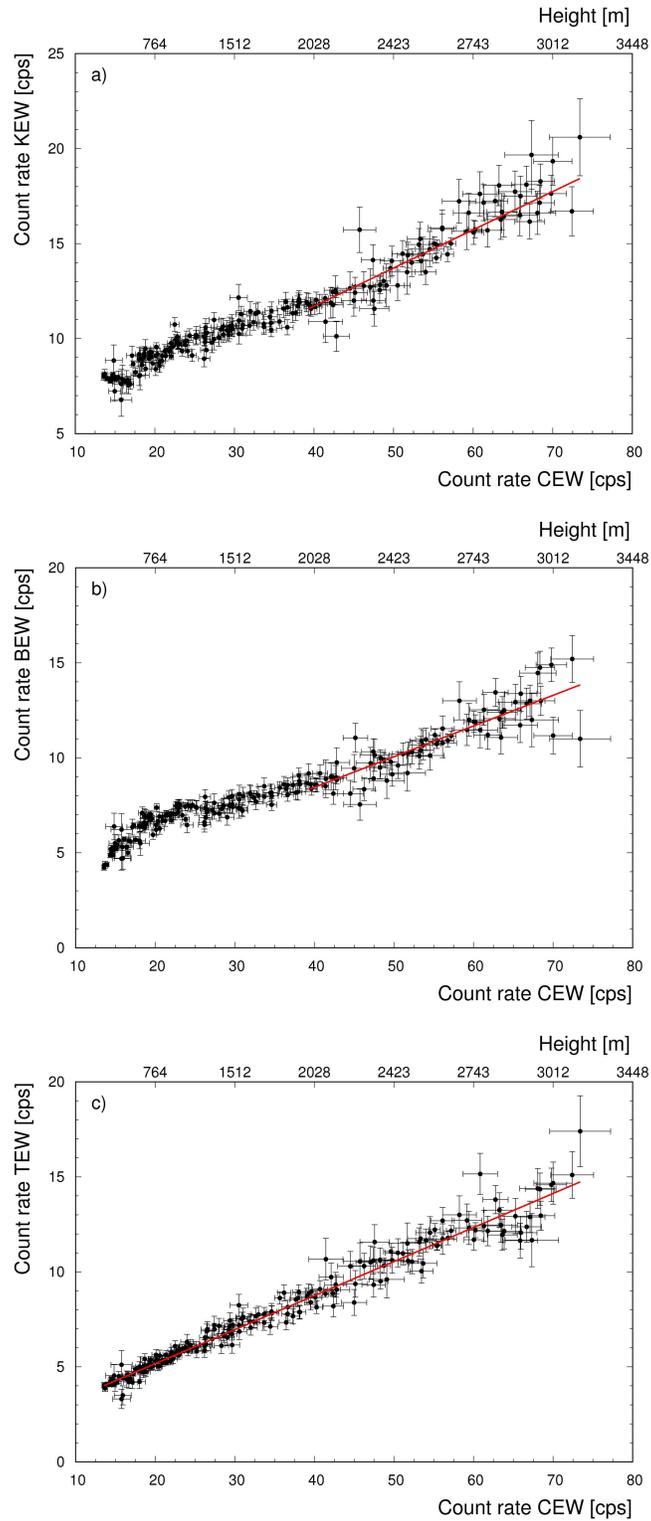

**Figure 7.** Panels a), b) and c) report respectively the experimental CRs (black points) in the KEW, in the BEW and in the TEW as function of the CR in the CEW, together with the corresponding fitting curve (solid red line).

Eq. 6 describes the sum of the constant aircraft CR plus the minimum cosmic CR component, corresponding to the one determined at zero altitude. As the right hand side of Eq. 6 can be calculated not only for the TEW, but also for the KEW and for the BEW, it is possible to estimate the minimum detectable CRs for the three energy windows of interest. These counting statistics can be naively converted to equivalent K, U and Th abundances homogeneously distributed across an infinite flat earth by means of sensitivity coefficients obtained from a dedicated ground calibration campaign on natural sites. According to this approach it is possible to estimate that the AGRS_16L detector can not measure K, U and Th concentrations lower than 0.05 $\cdot 10^{-2}$ g/g (15.7 Bq/kg), 0.4 $\mu$g/g (4.9 Bq/kg), 0.8 $\mu$g/g (3.2 Bq/kg), respectively.

In Fig. 8 the CED$^{EMS}$ calculated with the CARI-6P and EXPACS dosimetry tools shows an evident linear relation with the measured n$^{CEW}$ values. By fitting the scatter plots with:

$$CED^{EMS} = a_{CED^{EMS}} + b_{CED^{EMS}} n^{CEW} \tag{7}$$

an excellent (more than 0.99) $r^2$ coefficient of determination has been obtained in both cases. On the base of the $a_{CED^{EMS}}$ and $b_{CED^{EMS}}$ parameters reported in Fig. 8 caption, the AGRS_16L detector is calibrated for future measurements of CED$^{EMS}$. Although the described calibration method is clearly model dependent, the average discrepancy among CED$^{EMS}$ estimations is ∼10%, which is not so far from the typical uncertainties obtained with traditional methods.

For fixed detector and dosimetry tool, the slope and intercept parameters of Eq. 7 are not expected to vary significantly for different geomagnetic latitude and solar activity. On the other hand, the total CED comprises also a muon and a neutron component (respectively dominant at sea level and at high altitudes), together with additional minor contributions due to protons and He and heavy ions. In Appendix A we study the relation between CED and n$^{CEW}$ (see Fig. 9): in the temporal and spatial domain of our data taking, a linear relation between these two quantities is clearly observed for both CARI-6P and EXPACS calculations. Since the CED varies with geomagnetic latitudes and solar activities, the obtained linear curve parameters change for different data taking conditions. However, in the typical altitude range of AGRS surveys (z < 200 m), the maximum variation of the CED due to solar activity rarely exceeds 5%.

In Fig. 10 of Appendix A the ratio CED$^{EMS}$/CED is shown as function of the geographic latitude for four different altitudes, for a medium solar activity. As expected, the CED$^{EMS}$/CED ratio increases with increasing altitude, going from ∼14% at 0 m to ∼17% at 3000 m. A rule of thumb that can be formulated is that the ratio CED$^{EMS}$/CED∼0.15, where it has to be kept in mind that changing location, solar activity and dosimetry tool could bother this estimation.

## 5 Conclusion

This work illustrates the results of a ∼5 hour airborne offshore survey dedicated to the AGRS_16L detector calibration for the gamma background signal originating from cosmic radiation and equipment radioactivity and for the assessment of cosmic effective dose to human population. This airborne campaign has been conducted with the Radgyro, an ultra-light aircraft dedicated to multispectral airborne surveys, and has the peculiarity of having investigated a wide range of altitudes above sea level (77-3066 m). The acquisition of 17612 1 second spectra over the sea at different altitudes allowed to separate the background count rate into a constant aircraft component and a cosmic component exponentially increasing with increasing height.

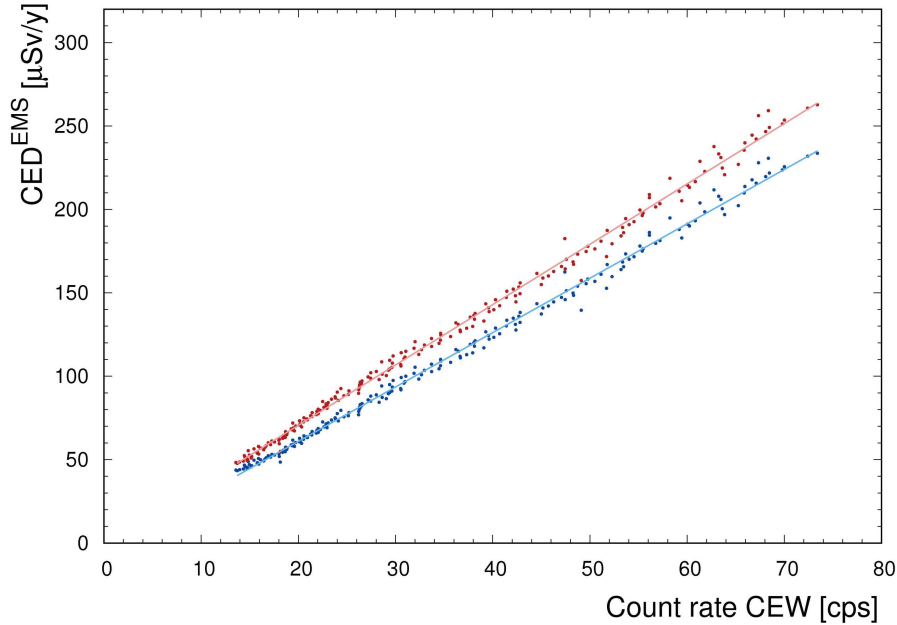

**Figure 8.** CED$^{EMS}$ obtained by running the CARI-6P (blue points) and the EXPACS (red points) softwares with fixed location (Viareggio, 43°56'N - 10°14'E) and fixed date (31 March 2016) corresponding to the data taking conditions versus the experimental CR in the CEW. The linear fitting curves (see Eq. 7) have best fit parameters equal to $a_{CED^{EMS}}$ = (-4.16 ± 0.59) $\mu$Sv/y and $b_{CED^{EMS}}$ = (3.26 ± 0.02) $\mu$Sv/(y·cps) for CARI-6P (light blue solid line) and $a_{CED^{EMS}}$ = (-1.67 ± 0.67) $\mu$Sv/y and $b_{CED^{EMS}}$ = (3.62 ± 0.02)$\mu$Sv/(y·cps) for EXPACS (light red solid line).

A statistical analysis has been performed to determine the parameters that linearly relate the count rate (CR) in the energy windows associated to the K, U and Th photopeaks and the counting statistics recorded in the cosmic energy window (CEW) in which no event coming from terrestrial radioactivity is expected. By monitoring the CR in the CEW and by applying the obtained linear relations it is possible to calculate for every airborne gamma-ray spectrum the background CRs in the photopeaks of interest that need to be subtracted prior the implementation of the height and stripping corrections before finally convert corrected elemental CRs to ground abundances. Minimum detectable K, U and Th abundances have been inferred from the minimum detectable CRs in the KEW, BEW and TEW, which correspond to the overall background CRs at zero altitude. On the basis of ground sensitivity coefficients, it is possible to assess that the minimum detectable abundances of the AGRS_16L detector are $0.05 \cdot 10^{-2}$ g/g, 0.4 $\mu$g/g, 0.8 $\mu$g/g, for K, U and Th respectively.

For the CRs in the CEW and in the TEW the exponential increase of counting statistics with respect to the altitude has been reconstructed, providing as argument for the exponential function a $\mu$ coefficient of $6 \cdot 10^{-4}$ m$^{-1}$ which is comparable with the values published in International Atomic Energy Agency (1991) and Grasty and Minty (1995). Moreover, the analysis of the CRs in the TEW highlighted a perfect internal consistency among linear fit and exponential fit parameters. The exponential analysis for the CRs in the KEW and in the BEW was unfeasible due to the application of a low altitude cut to the dataset (z>2000 m), which allowed to exclude potential contamination caused by atmospheric $^{222}$Rn. This point, however, deserves a deeper investigation as deviations from purely exponential/linear behaviors could in principle be used to quantify the atmospheric $^{222}$Rn abundance at different elevations (Baldoncini et al., 2017).

The AGRS_16L has also been calibrated for assessing the electromagnetic shower component of the cosmic effective dose (CED$^{EMS}$) to human population by using as calibrating reference the dose rate values obtained separately with the CARI-6P and EXPACS softwares. The relation between the CR in the CEW and the CED$^{EMS}$ has been found to be linear. Although this approach for calibrating an AGRS detector for CED$^{EMS}$ is clearly model dependent, the results are in agreement at ∼10% level. This quality of this estimation is comparable with traditional approaches. Finally, we observed a good linear relation between the cosmic effective dose (CED) and the count rate in the CEW (Fig. 9) as well as an almost constant profile of the CED$^{EMS}$/CED ratios at different latitudes of about 15% for typical AGRS survey altitudes.

## Appendix A

The purpose of this section is to investigate the possibility of inferring the cosmic effective dose starting from a direct count rate measurement performed with an AGRS detector. In Fig. 9 we report the CED, calculated with the CARI-6P and EXPACS dosimetry softwares as function of the measured $n^{CEW}$, together with the linear fitting curves defined according to the following equation:

$$CED = a_{CED} + b_{CED} n^{CEW} \tag{A1}$$

An excellent linear relation between CED and $n^{CEW}$ characterized by a $r^2$ coefficient of determination greater than 0.99 is observed for both dosimetry tools.

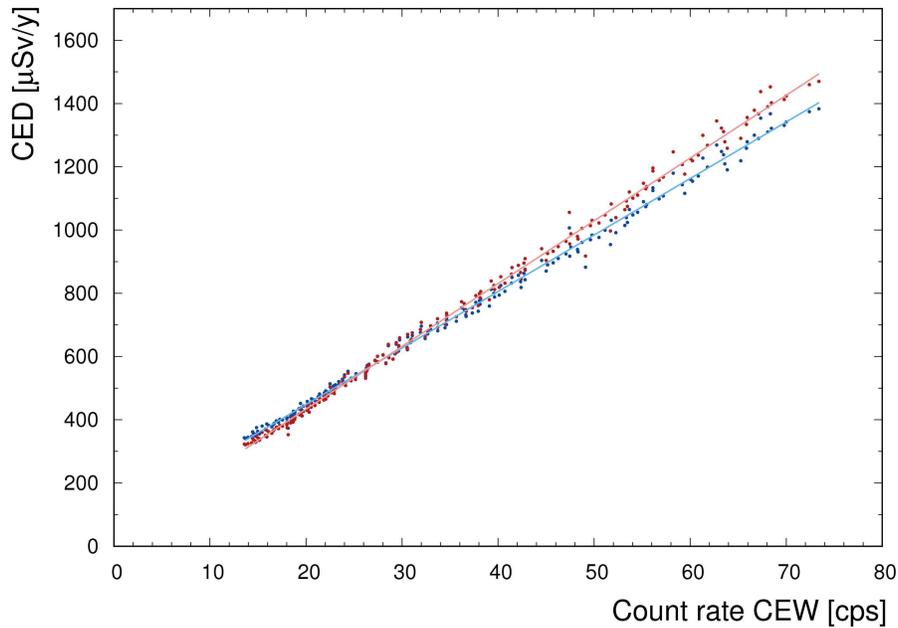

**Figure 9.** CED obtained by running the CARI-6P (blue points) and the EXPACS (red points) softwares with fixed location (Viareggio, 43°56'N - 10°14'E) and fixed date (31 March 2016) corresponding to the data taking conditions versus the experimental CR in the CEW. The linear fitting curves (see Eq. A1) have best fit parameters equal to $a_{CED}$ = (90.9 ± 3.1) µSv/y and $b_{CED}$ = (17.9 ± 0.1) µSv/(y·cps) for CARI-6P (light blue solid line) and $a_{CED}$ = (36.6 ± 3.4) µSv/y and $b_{CED}$ = (19.9 ± 0.1) µSv/(y·cps) for EXPACS (light red solid line).

With the purpose of testing how a change of latitude in AGRS surveys could affect the CED estimation, we reconstruct the $CED^{EMS}$/CED ratios along a meridian at different altitudes. In Fig. 10 we show the $CED^{EMS}$/CED ratios calculated with the CARI-6P and EXPACS dosimetry softwares as function of the geographic latitudes in the (0 - 3000) m range. In both cases it is possible to observe that the ratio generally increases for increasing altitude and that it reaches a plateau for latitudes greater than 50°. For varying solar activities, the calculated $CED^{EMS}$/CED profiles follow the same trends with a negligible variation with respect to the medium solar activity scenario of Fig. 10. Finally, as the $CED^{EMS}$/CED profile is reasonably smooth in the typical AGRS altitude range (z < 200 m), this evidence adds a point in favor of the presented method for the estimation of the CED by using direct gamma-ray measurements.

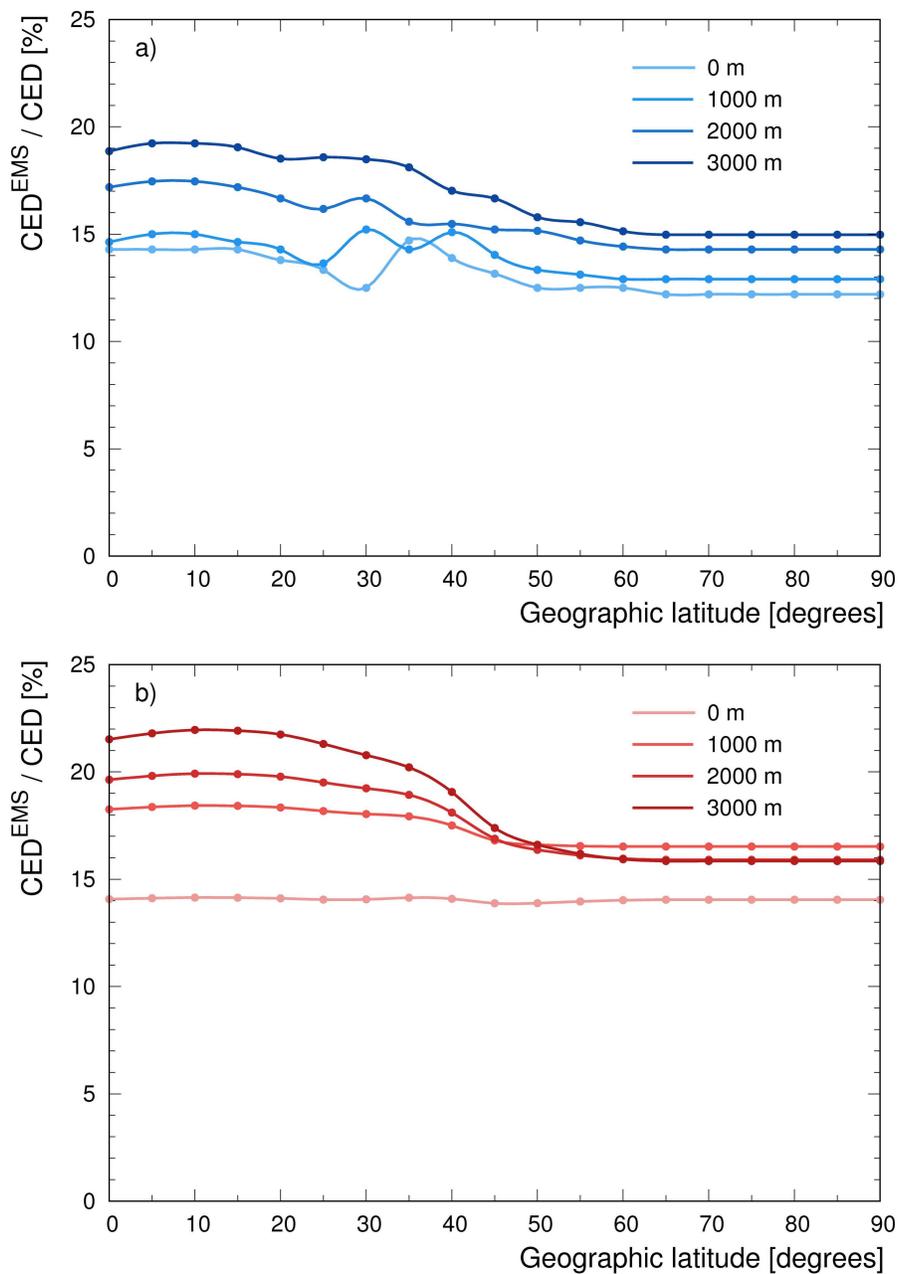

**Figure 10.** $CED^{EMS}/CED$ ratios as function of the geographic latitude calculated for a medium solar activity (31 March 2016) and for four different altitudes (0 m, 1000 m, 2000 m and 3000 m) by using the CARI-6P (panel a) and the EXPACS (panel b) dosimetry tools.

*Acknowledgements.* This work was partially founded by the National Institute of Nuclear Physics (INFN) through the ITALian RADioactivity project (ITALRAD) and by the Theoretical Astroparticle Physics (TAsP) research network. The co-authors would like to acknowledge the support of the Geological and Seismic Survey of the Umbria Region (UMBRIARAD), of the University of Ferrara (Fondo di Ateneo per la Ricerca scientifica FAR 2016), of the Project Agroalimentare Idrointelligente CUP D92I16000030009 and of the MIUR (Ministero dell'Istruzione, dell'Universitá e della Ricerca) under MIUR-PRIN-2012 project.

The authors are grateful to Kyle Copeland for providing the CARI-6 software versions. The authors would like to thank also the staff of GeoExplorer Impresa Sociale s.r.l. for their support and Mauro Antogiovanni, Claudio Pagotto, Gianluca Tomasi, Ivan Callegari, Gerti Xhixha, Merita Kaçeli Xhixha and Andrea Motti for their collaboration which made possible the realization of this study. The authors would also like to show their gratitude to Enrico Chiarelli, Gianni Fiorentini, Eligio Lisi, Michele Montuschi, Barbara Ricci and Carlos Rossi Alvarez for useful comments and discussions. The comments from the anonymous reviewers improved the quality and clarity of our manuscript.